\documentclass[aps,prd,eqsecnum,nofootinbib,superscriptaddress,preprintnumbers,11pt]{revtex4}

\usepackage{amssymb,amsmath,psfrag,epsfig,feynarts}

\newcommand{\ud}{\mathrm d}
\newcommand{\dsi}{\frac{\ud s_i}{s_i}\frac{\ud \bar s_i}{\bar s_i} s_i^{h+c_i} \bar s_i^{h-c_i}}
\newcommand{\pn}{$\phi^n$}
\newcommand{\iads}{\int_{\mbox{\tiny $AdS$}}}
\newcommand{\ib}{\int_{\partial\mbox{\tiny $AdS$}}}

\renewcommand{\eqref}[1]{(\ref{#1})}

\newcommand{\gt}[1]{\Gamma\left(\frac{ #1}{2}\right)}

\newcommand{\dd}[1]{\Delta_{#1}}
\newcommand{\bs}[1]{\bar s_{#1}}

\newcommand{\h}{h}

\newcommand{\D}{\mathcal{D}}
\newcommand{\Y}{\mathcal{Y}}

\newcommand{\Aone}{s_1 B_1}
\newcommand{\Atwo}{s_2 B_2 + \bs2(\bs1\Aone )}

\renewcommand{\O}{\mathcal{O}}

\newcommand{\M}{\mathcal{M}}
\newcommand{\Bz}{\mathcal{B}^Z}

\newcommand{\be}{\begin{eqnarray}}
\newcommand{\ee}{\end{eqnarray}}

\newcommand{\beq}{\begin{equation}}
\newcommand{\eeq}{\end{equation}}
\newcommand{\beqa}{\begin{eqnarray}}
\newcommand{\eeqa}{\end{eqnarray}}
\newcommand{\bea}{\begin{aligned}}
\newcommand{\eea}{\end{aligned}}
\newcommand{\bem}{\begin{multline}}
\newcommand{\eem}{\end{multline}}
\newcommand{\nn}{\nonumber  \\}

\newcommand{\Gd}{\delta}
\newcommand{\GD}{\Delta}

\def\ket(#1,#2){\langle #1\,#2\rangle}
\def\mt(#1,#2,#3,#4){\langle #1\,#2\,#3\,#4\rangle}

\begin{document}

\begin{center}%
{\Large\textbf{\mathversion{bold}%
On Feynman rules for Mellin amplitudes in AdS/CFT}}

\author{Dhritiman Nandan}
\email{dhritiman_nandan@brown.edu}
\affiliation{Physics Department, Brown University, Providence, Rhode
Island 02912, USA}

\author{Anastasia Volovich}
\email{anastasia_volovich@brown.edu}
\affiliation{Physics Department, Brown University, Providence, Rhode
Island 02912, USA}

\author{Congkao Wen}
\email{c.wen@qmul.ac.uk }
\affiliation{Centre for Research in String Theory, School of Physics and Astronomy,\\
Queen Mary University of London, Mile End Road, London, E1 4NS, United Kingdom.}

\maketitle

\vspace{3cm} \textbf{Abstract}\vspace{7mm}

\begin{minipage}{12.7cm}
The computation of CFT correlation functions via Witten diagrams in AdS space can be
simplified via the Mellin transform. Recently a set of Feynman rules for tree-level Mellin space amplitudes has been proposed for scalar theories. In this note we derive these rules
by explicitly evaluating all of the relevant Witten diagram integrals for the scalar $\phi^n$ theory. 
We also check that the rules reduce to the usual Feynman rules in the flat space limit.

\end{minipage}

\end{center}

\newpage
\section{Introduction}

AdS/CFT is a powerful tool \cite{Maldacena:1997re,Witten:1998qj,Gubser:1998bc} which among other things allows us to compute CFT correlators at strong coupling via Witten
diagrams in AdS space. In practice these computations are still quite challenging in position space and generally require
a lot of work, see \cite{LiuTseytlin,hep-th/9811152,D'Hoker:1998mz,Freedman:1998bj,Freedman:1998tz,D'Hoker,Arutyunov:2000py,Arutyunov:2002fh,Arutyunov:2003ae,Berdichevsky:2007xd,Uruchurtu:2008kp,Buchbinder:2010vw,Uruchurtu:2011wh,Dolan:2006ec,Howtozintegrals, Raju1,Raju2}.
Recently it has been argued that taking the Mellin transform of correlation functions 
drastically simplifies the computations and the resulting expressions have
nice mathematical structure, see e.g.~\cite{Mack:2009mi,Mack:2009gy,Penedones:2010ue,Paulos:2011ie,Fitzpatrick:2011ia,
Costa:2011mg,Costa:2011dw,arXiv:1102.0577}.
 The correlation functions of primary scalar operators for a CFT can be written in Mellin space as 
 \beqa
 \langle \O(x_1)\O(x_2)\ldots\O(x_n)\rangle \sim
 \int \ud \delta_{ij} M(\Gd_{ij})\! \prod_{1\leq i<j \leq n}\!\!\Gamma(\delta_{ij})\,(x^2_{ij})^{-\delta_{ij}} ,
 \label{MellinMack}
\eeqa 
 where  $ M(\Gd_{ij} )$ is called the Mellin amplitude and parameters $\delta_{ij}$ can be parametrized as 
 $ \delta_{ij} = k_i \cdot k_j $. Mellin amplitudes have many similarities to scattering amplitudes in flat space, in particular the large AdS radius limit of the Mellin amplitude was argued to be equivalent to the scattering amplitudes in flat space \cite{Penedones:2010ue,Fitzpatrick:2011ia}, such that $ k_i $ plays the role of momentum in flat space,\footnote{The quantities $k_i$'s are also called ``Mellin momentum" or ``fictitious momentum" but for the sake of brevity we will just refer to them as ``momentum" in the rest of the paper and we stress that this interpretation is precise only when we look at the flat space limit of the Mellin amplitude and not in general. } suggesting that Mellin amplitudes could be used to provide a holographic definition of the S-matrix, see \cite{susskind,polchinski,GGP,JP,Katz,TakuyaFSL,Fitzpatrick:2011jn,Gary:2009mi,GiddingsBulkLoc,Gary:2011kk} for related discussion.
  
More recently in~\cite{Paulos:2011ie} and \cite{Fitzpatrick:2011ia}, the authors studied various aspects of the Mellin representation of AdS correlators. In particular, 
a set of Feynman rules, for computing Mellin amplitudes for any theory of scalar field at tree-level, was proposed
and checked for a few non-trivial correlators in $\phi^3$ and $\phi^4$ theories in~\cite{Paulos:2011ie}
as well as
recursively via a factorization formula for $\phi^3$ theory in~\cite{Fitzpatrick:2011ia}.

In this note we will consider a scalar field with $\phi^n $ interaction 
at tree level and offer a direct proof of the Feynman rules for Mellin amplitudes  by evaluating all the Witten diagram integrals explicitly. 
We hope that our results will be useful for better understanding of the structure of Mellin amplitudes and for
the future development of similar rules for fields with spin and for loop amplitudes.
 We have also checked that the  Mellin space Feynman rules reduce to the usual Feynman rules in the flat space limit.

The paper is organized as follows. In section $ 2 $ we review the conjectured Feynman rules for Mellin amplitudes. In section $ 3 $ we use particular forms of the bulk-to-boundary and bulk-to-bulk propagators to compute the Witten diagram with the maximal off-shell vertex for a $ \phi^n $ theory and show that they lead to the conjectured Feynman rules for Mellin amplitudes. Then we demonstrate that we get the same Feynman rules for such a vertex embedded in a very general Witten diagram of the $ \phi^n $ theory. In section $ 4 $ we show that these Feynman rules reduce to the usual Feynman rules in the flat space limit. 

\smallskip

{\bf Note added:} While this paper was in preparation, the paper~\cite{Fitzpatricknew} appeared which checks the formula for 
the off-shell $n$-pt vertex of the $\phi^n$ theory via recursion relations.
 
 \section{Feynman rules for Mellin amplitudes}

Let us first review the Feynman rules for Mellin amplitudes corresponding to any tree level Witten diagram in AdS$_{d+1}$ for a $ \phi^n $ scalar theory, as proposed in \cite{Paulos:2011ie}. To compute the Mellin amplitude one has to put together propagators and
vertices following a few simple steps:

 $-$ Assign a ``momentum" $k_i$ to every line such that  the external lines of the Witten diagram have  $-k_i^2=\Delta_i$ and at each vertex we have conservation $\sum_i k_i=0$,\footnote{ The vector $k$ has such properties because it solves the constrains of $\delta_{ij}$, namely $\delta_{ij} = \delta_{ji}$ and $\sum_{j \neq i} \delta_{ij} = -\Delta^2_i$.} where $\Delta_i$ is the conformal dimension of the corresponding field. 
 
 $-$ 
 Assign an  integer $m_i$ to each internal line with the  propagator 
\beq
	{\cal P}_i=\frac{-1}{2 m_i! \Gamma(1+\Delta_{m_i}+m_i-h)}\
	\frac{1}{k_i^2+(\Delta_{m_i} +2 m_i)},
\label{mellinprop}
\eeq
where $h=d/2.$

\begin{figure}[h]
\begin{center} 
\includegraphics[width=9cm]{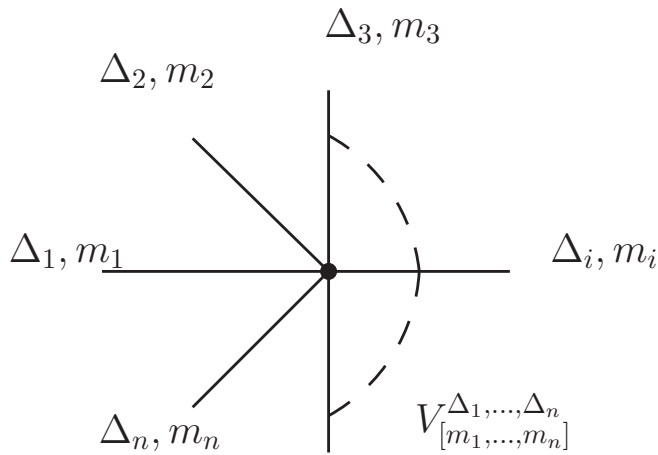}
\caption{\small{A general vertex for $ \phi^n $ theory.}} 
\label{figvertex}\end{center}

 \end{figure}

$-$ The factor for a vertex connecting lines with dimension $\Delta_i$ and integers $m_i$, (see Fig.~\ref{figvertex}) is given by
	\beqa \label{vertexgeneral}
 && V^{\dd 1\ldots \dd n}_{[m_1,\ldots, m_n]}=g^{(n)}\, \gt{\sum_{i=1}^n \dd i-2h}
	\left(\prod_{i=1}^n 	\left(1-h+\dd i\right)_{m_i}\right)\nonumber \\
 && F_A^{(n)}\left(\frac {\sum_{i=1}^n \dd i-\!2h}2,\left\{-\!m_1,\ldots,-\!m_n\right\},\left\{1\!+\!\dd1\!-\!h,\ldots,1\!+\!\dd n\!-\!h\right\};1,\ldots,1\right) \hspace{1 cm} \ \ \  
	\eeqa
	where  $g^{(n)}$ is the coupling in the $g^{(n)} \phi^n$ theory, $(a)_m = {\Gamma(a+m) \over \Gamma(a)}$ is the Pochhammer symbol and $F_A^{(n)}$ is the Lauricella function of $n$ variables 
	\be
F_A^{(n)}\left(y,\left\{a_1,\ldots,a_n\right\},\left\{b_1,\ldots,b_n\right\};x_1,\ldots,x_n\right) &=&
\sum_{l_i=0}^{\infty}\left( (y)_{\sum_{i=1}^n l_i} \prod_{i=1}^n\frac{ (a_i)_{l_i}}{(b_i)_{l_i}} \frac{x_i^{l_i}}{l_i!}\right). 
\label{lauricella}
\ee

$-$ Finally, sum over all positive integers $m_i$ to obtain the Mellin amplitude. 

We note here that the vertex given above is the most general type of vertex (or the maximal off-shell vertex) when all legs are off-shell\footnote{The legs connecting to the AdS boundary directly are referred to as the on-shell legs, while those that do not connect to the boundary are the off-shell legs.}, but the theory would also have vertices with less number of off-shell legs.  The vertex factor in such cases can be simply obtained from the general case by taking some of the $ m_i $'s to zero, corresponding to the legs going on-shell. 
Also, note that the Lauricella function of $ m $
 variables can be written in a series form as in (\ref{lauricella}) which is convergent for $ \sum_i |x_i|<1 $. For the vertex above, all $ n $ variables $x_i$ take a particular value $ 1 $, which is the Lauricella function evaluated at that particular point, which is well-defined via analytic continuation.

\section{Proof of Feynman rules }
\subsection{Maximal off-shell vertex}

     \begin{figure}[h]
\begin{center} 
\includegraphics[width=9cm]{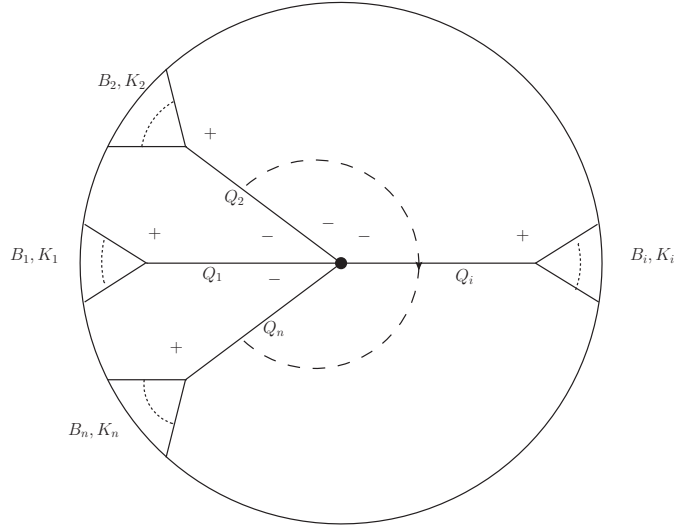}
\caption{\small{The Witten diagram for the scalar $\phi^n$ theory with a vertex having $n$ off-shell legs and $n$ vertices having $1$ off-shell leg and $(n\!-\!1)$ on-shell legs.}} 
\label{figphin}\end{center}

 \end{figure}

In this section we consider a Witten diagram for the scalar $\phi^n$ theory in AdS$_{d+1} $ which has a vertex with the maximal number of off-shell legs (see Fig.~\ref{figphin}) and prove the Feynman rules for this case which we described in the previous section.

Let $X^I$ be the coordinates of the Euclidean  AdS$_{d+1} $ space, 
embedded in a $ d+2 $ dimensional Minkowski space such that $ X^2 =- R^2, $ where $ R $ is the AdS radius and the point on the boundary $P^A$ defined on the light-cone such that $P^2=0$.
The bulk-to-boundary propagator between a point $ P $ on the boundary and $ X $ in the bulk for a scalar field of dimension $ \GD$  is given by\footnote{We will drop the normalization factor $ \frac{1}{2\pi^h \Gamma(1+\Delta-h)} $ from \eqref{db} for subsequent calculations, since it is not relevant. Moreover this factor goes into the overall normalization factor in the definition of Mellin amplitudes and according to \eqref{MellinMack} we have ignored it in this note and the inclusion of this factor would allow us to write \eqref{MellinMack} as an equality. }
	\be
		E(P,X)=	\frac{1}{2\pi^h \Gamma(1+\Delta-h)}
		\int_0^{+\infty} \frac{\ud t}{t} t^{\Delta}\, 
		e^{2 t P\cdot X}. \label{db}
	\ee
The bulk-to-bulk propagator between the points $X_1$ and $X_2$ can be written as 
an integral over a point $Q$ on the boundary of the AdS, the integrand being the product of two bulk-to-boundary propagators 
of states with unphysical dimension $h \pm c$
\beq
		G_{BB}(X_1,X_2) = 
		\int_{-i \infty}^{+i \infty}\frac{\ud c}{2\pi i} f_{\Delta}(c)\ib \ud Q\int \frac{\ud s}{s}\frac{\ud \bar s}{\bar s} s^{h+c} \bar s^{h-c}\, 
		e^{2 s Q\cdot X_1+2 \bar s Q\cdot X_2}  
		\label{bb}
	\eeq 
where 
	\be
	f_{\Delta}(c)\equiv \frac{1}{2\pi^{2h}[(\Delta-\h)^2-c^2]}	\frac{1}{\Gamma(c)\Gamma(-c)}.
	\ee

To simplify the notations, let us call each set of external $(n\!-\!1)$ legs in Fig.~\ref{figphin} as a block, and denote it as $B_i$ with $i=1, \dots, n$. 
Equations (\ref{bb}) and (\ref{db}) give us the building blocks of any arbitrary Witten diagram in the $ \phi^n $ theory. Fig.~\ref{figphin} can be constructed using two types of $n$-point correlation functions, built out of \eqref{db} and \eqref{bb}, which are given as
\beqa \label{subamp}
		A_n(B_i,Q_{i,+})&=&g^{(n)}\, \int_0^{+\infty} 
		\prod_{i=1}^{n-1}\frac{\ud t_i}{t_i}t_i^{\Delta_i} \frac{\ud s_i}{s_i} s_i^{\h+c_i} 
		\iads \ud X_i\, e^{2(B_i+s_i Q_i)\cdot X_i},\\ \nonumber
		A_n(Q_{1,-},\ldots, Q_{n,-})&=&g^{(n)}\, \int_0^{+\infty} 
		\prod_{i=1}^{n}\frac{\ud \bar s_i}{\bar s_i}\bar s_i^{\h-c_i} 
		\iads \ud X_i\, e^{2(\sum_{k=1}^n\bar s_k Q_k)\cdot X_i},
	\eeqa
	where the blocks of $ (n\!-\!1) $ legs, which we call $ B_i $, are typically given as,
	\beq
	B_i=\sum_{k=(i-1)(n-1)+1}^{i(n-1)}t_kP_k.
		\label{block}
	\eeq
	We note that in general $ B_i $ can also contain fewer legs, (in which case the Witten diagram gives a vertex with fewer off-shell legs) so the limits of the summation in (\ref{block}) would change according to the diagram under consideration. Moreover in Fig.~\ref{figphin} the label $ K_i $ indicates
	\beq
	 K_i \equiv k_{a_i}+\ldots + k_{a_i+n-1}
	 \eeq
	the sum of the momenta of all the fields in the block $ B_i $ where $ k_{a_i} $ is the momenta of each field.
	
Now, let us write the expression for the Witten diagram in Fig.~\ref{figphin} using the $ n$-point correlation functions (\ref{subamp}) as the building blocks and we get
\beqa
		 A&=& \int_{-i\infty}^{+i \infty} \prod_{i=1}^{n}\frac{\ud c_i }{(2\pi i)^n} f_{\Delta_i}(c_i)\ib 	\prod_{i=1}^n\ud Q_i \biggl(A_n(B_i,Q_{i,+})\ldots A_n(B_n,Q_{n,+}) A_n(Q_{1,-},\ldots, Q_{n,-})\biggr)\nn
		 &=& (g^{(n)})^{n+1}  \int_{-i\infty}^{+i \infty} \prod_{i=1}^{n}\frac{\ud c_i }{(2\pi i)^n} f_{\Delta_i}(c_i)\int \prod_{i=1}^{n(n-1)} \frac{\ud t_i}{t_i} t_i^{\Delta_i} 
		\int\prod_{i=1}^{n}\dsi \nn &\times & \ib 	\prod_{i=1}^n\ud Q_i  \iads \prod_{i=1}^{n+1}\ud X_i 
	\exp \biggl((2 \sum_{i=1}^nX_i\cdot(B_i +s_i Q_i)+2 X_{n+1}\cdot(\sum_{i=1}^n\bar s_i Q_i)\biggr),
	 \label{anpoint}
\eeqa
where $A$ is the $n(n-1)$ point correlation function. 
\subsection{Evaluating the integrals}
\subsubsection*{Integrals over $ X_i $}
The integrations over the bulk points $ X_i $ can be done by applying (\ref{Xint}) from the Appendix, which gives us the result,
\beqa
 A =(g^{(n)})^{n+1} (\pi^h)^{n+1} \int_{-i\infty}^{+i \infty} \prod_{i=1}^{n}\frac{\ud c_i }{(2\pi i)^n} f_{\Delta_i}(c_i)\int \prod_{i=1}^{n(n-1)} \frac{\ud t_i}{t_i} t_i^{\Delta_i} 
		\int\prod_{i=1}^{n}\dsi&& \nonumber \\
	\ib 	\prod_{i=1}^n\ud Q_i \biggl(\prod_{i=1}^n \Gamma(\frac{ \Delta_{B_i}+(h+c_i)-2h}{2})\biggr) \Gamma(\frac{\sum_{i=1}^n(h-c_i)-2h}{2}) e^{E_Q},
\label{anpointX}	
	\eeqa
where $ \Delta_{B_i} = \sum_{j \in B_i} \Delta_j$ and $\Delta_j$ is the conformal dimension of the field $j$ and the exponent in the above integrand is given by,
\beqa
E_Q&=&\sum_{i=1}^{n}(B_{i} +s_{i} Q_{i})^2+(\sum_{i=1}^{n} \bar{s_i}Q_i)^2.
\label{En0}
\eeqa

\subsubsection*{Integrals over $ Q $}
To perform the $ Q_i $ integrals we first expand \eqref{En0}  and rewrite it as,
\beqa
 E_Q    &=& \sum_{i=1}^{n}\bigl(B_i^2+2s_iQ_i\cdot B_i\bigr)+2 \sum_{\substack{1\leq i<j \leq n}}\bar s_i\bar s_j Q_i\cdot Q_j.
 \label{En}
\eeqa
Now we will integrate out the $ Q_i$'s successively.

First we will do the $ Q_1 $ integral using (\ref{boundint}) and  using the on-shell condition, $ Q_i^2=0 $ to simplify the result, we finally get,\footnote{ At each step of doing the $ Q $ integral we will get a $ 2\pi^h $ factor which we would drop in the following steps to help us reduce clutter. }
\beqa
E_{Q_1}&=&\sum_{i=1}^n B_i^2+ (\Aone)^2\nn
&+& 2\biggl((1+\bs1^2 )(\sum_{\substack{1<i<j}}^n
\bar s_i\bar s_j Q_i \cdot Q_j)+\sum_{i\neq 1}^{n}( s_i B_i +\bar s_i (\bs1 \Aone))\cdot Q_i\biggr),
\label{en1}
\eeqa
where the notation $ E_{Q_k} $ is used to denote the exponent obtained as a result of doing the set of successive integrals from $ Q_1 $ to $ Q_k $, i.e.
 \beq
  \ib 	\prod_{i=1}^k\ud Q_i e^{ E_Q} = e^{E_{Q_k}}.
  \nonumber
  \eeq

Next, we perform  the $ Q_2 $ integral in a similar way and we find that $E_{Q_2}$ can be written as
\beqa
 E_{Q_2}&=&\sum_{i=1}^n B_i^2+ (\Aone)^2+(\Atwo)^2+ 2\biggl((1+\bs1^2 )(1+\bs2^2(1+\bs1^2))(\sum_{\substack{2<i<j}}^n
\bar s_i\bar s_j Q_i \cdot Q_j)\biggr)\nn
&+& 2\biggl(\sum_{i\neq \{1,2\}}^{n}\biggl( s_iB_i +\bar s_i \bigl((\bs1 \Aone)+ (1+\bs1^2)\bs2 (\Atwo)\bigr)\biggr)\cdot Q_i\biggr).
\label{en2}
\eeqa

We continue integrating out the $ Q_i $'s successively as in the last few steps and  integrating the $ p_{th} $ step the result is of the form,
\beqa
E_{Q_p}&=& \sum_{i=1}^n B_i^2+ \sum_{i=1}^p \big( \bar{s}_i Y_{i-1} + s_i B_i \big)^2+ 2\biggl((\prod^{p}_{m=1}g_m)(\sum_{\substack{p<i<j}}^n
\bar s_i\bar s_j Q_i \cdot Q_j)\biggr)\nn
&+& 2\biggl(\sum_{i\neq \{1,2,\ldots, p\}}^{n}( s_i B_i +\bar s_i Y_p)\cdot Q_i\biggr),
\label{enp}
\eeqa
where we have defined the functions $ Y_i $ and $ g_i $ as,
 \be \label{recursion}
Y_l&=&\sum^l_{i=1} {(\prod_{k=1}^{l}g_k)\over g_i} s_i\bar{s}_iB_i\quad \rm{and}\\ \nonumber
g_l&=&(1+\bar{s}^2_l \prod^{l-1}_{k=0}g_{k}) ~~~ {\rm with} ~~g_0=1.
\ee


After integrating out all the $ Q_i $'s using (\ref{enp}) we finally get the exponent of the integrand in \eqref{anpointX} as,
\be 
E_{Q_n}=\sum_{i=1}^{n}B_i^2 + \sum^n_{l=1}\big( \bar{s}_l Y_{l-1}+ s_l B_l \big)^2.
\label{phin1}
\ee
 \subsubsection*{Integrals over $ t_i $}
Let us first expand the term in the parentheses in (\ref{phin1}), and with the help of \eqref{recursion} we get

\beqa
E_{Q_n} &=&\sum_{i=1}^{n}(1+s_i^2 F_i)B_i^2+\sum_{1\leq i <j\leq n}\frac{2 (s_i \bar s_is_j \bar s_j)(B_i\cdot B_j)}{g_ig_j}\bigl(\prod_{l=1}^n g_l\bigr).
\label{Aexpand}
\eeqa
where
\beqa
F_i=1+{\bar{s}^2_i  \over g^2_i} (\sum^n_{l={i+1}} \bar{s}^2_l(\prod_{k=1}^{l-1}g_k^2)).
\label{Aexpand0}
\eeqa
Next, we will apply Symanzik star formula (\ref{symanzikint}) to our integral (\ref{anpointX}) in order to obtain the Mellin amplitude $M(\delta_{ij})$.

Let us recall that $ B_i $'s are given as $\sum t_lP_l $, so the exponent of the integrand would only have terms quadratic in $ t $
coming from expanding the $ B_i^2 $ and $ B_i\cdot B_j $ terms in (\ref{Aexpand})\footnote{In the embedding formalism, $ P_{ab} =-2 P_a \cdot P_b$ and $ P_a^2=0 $}.
Using (\ref{symanzikint}), we can see that the full result of the Witten diagram integral gives,
\beqa
 A &=&(g^{(n)})^{n+1}
		\frac{(\pi^h)^{n+2} }{2(2\pi i)^{\frac 12 n(n-3)}}\int \ud \delta_{ij}\! \prod_{1\leq i<j \leq n}\!\!\Gamma(\delta_{ij})\,(P_{ij})^{-\delta_{ij}} \int_{-i\infty}^{+i \infty} \prod_{i=1}^{n}\frac{\ud c_i }{(2\pi i)^n}\nn
		&\times &\biggl( \bigl( \prod_{i=1}^{n}\Gamma(\frac{ \Delta_{B_i}+(h+c_i)-2h}{2})\bigr) \Gamma(\frac{\sum_{i=1}^n(h-c_i)-2h}{2})f_{\Delta_i}(c_i) \M(k_i,c_i)\biggr),
\label{Waftertint}	
	\eeqa 
	where we introduce a new notation $\M(k_i,c_i)$ and call it as the Mellin integrand which is given as
\be \label{phin2}
\M(k_i,c_i)&=&\int \prod^n_{i=1} {ds_i \over s_i} {d\bar{s}_i \over \bar{s}_i} \, s_i^{h+c_i+a_i} \bar{s}^{h-c_i+a_i}_i (g_i)^{b_i} \big( 1+s^2_i F_i \big)^{d_{i}},
\ee
such that the Mellin amplitude can be given in terms of the Mellin integrand as,
\beqa
M(\Gd_{ij})&=&\int_{-i\infty}^{+i \infty} \prod_{i=1}^{n}\frac{\ud c_i }{(2\pi i)^n}\bigl( \prod_{i=1}^{n}\Gamma(\frac{ \Delta_{B_i}+(h+c_i)-2h}{2})\bigr)\nn
		&\times &  \Gamma(\frac{\sum_{i=1}^n(h-c_i)-2h}{2})f_{\Delta_i}(c_i) \M(k_i,c_i).
\eeqa
Note that in the Mellin integrand $\M(k_i,c_i)$ we have used $k_i$ instead of $\delta_{ij}$, and recall that $\delta_{ij} \equiv k_i \cdot k_j $.

Furthermore, a few words about the exponents $ a_i, b_i$ and $d_i $ are in order. With respect to the $i_{th}$ propagator in Fig.~\ref{figphin}, $ a_i $ is the product of the momenta flowing through the propagator $ i $ from both sides. So according to our convention of Fig.~\ref{figphin}, where $ K_i $ is the sum of all momenta of the fields contained in the block $ B_i $, i.e. $K_i \equiv k_{a_i}+\ldots+ k_{a_i+n-1}$, we get 
\beqa
   a_i \equiv -(K_i \cdot \sum_{m \neq i}K_m) = K^2_i.
   \label{aidef}
  \eeqa
   The exponent  $b_i$ is the sum of all possible products of the momenta flowing through the propagators connecting the propagator $i$ on the $\bar s_i$ side i.e.
   \beqa 
  b_i \equiv -(\sum_{m,n\neq i , m\neq n}K_m\cdot K_n ), \label{bidef}
   \eeqa
   while $d_i$ is the sum of all possible products of the momenta flowing from the other direction, namely $ s_i $ side,
   \beqa
      d_i \equiv  -{a_i - \Delta_{B_i} \over 2} ,
      \label{didef}
  \eeqa
       recall that $ \Delta_{B_i} = \sum_{j \in B_i} \Delta_j$.

\subsubsection*{Integrals over $ s_i $ and $ \bar s_i $}

The integral $ \M(k_i,c_i) $ in (\ref{phin2}) can be greatly simplified using a set of transformations which are the generalization of the transformations used in \cite{Paulos:2011ie}. Firstly, we rescale $s^2_i$ by a factor of $F_i$, then (\ref{phin2}) becomes
\be \label{phin3}
\M(k_i,c_i)&=&\prod^n_{i=1} \int {ds_i \over s_i} s^{h+c_i+a_i}_i ( 1+s^2_i )^{d_{i}} \int {d\bar{s}_i \over \bar{s}_i} \, F_i^{-{h+c_i+a_i \over 2}} \bar{s}^{h-c_i+a_i}_i g_i^{b_i}.
\ee
Then we can make a set of consecutive transformations on $\bar s$'s, to simplify the integral further,
\be
\bar{s}^2_j &\rightarrow &x_j \\ \nonumber
x_2 &\rightarrow &{x_2 \over 1+x_1} \\ \nonumber
&\vdots &\\ \nonumber
x_n &\rightarrow &{x_n \over (1+x_1)(1+x_2)\dots(1+x_{n-1})} \\ \nonumber
x_1 &\rightarrow &{x_1 \over (1+x_2)(1+x_3)\dots(1+x_n)} \\ \nonumber
&\vdots &\\ \nonumber
x_{n-1} &\rightarrow &{x_{n-1} \over 1+x_n}.
\label{transformation}
\ee
Under the above set of transformations we find that $g_i$, $F_i$ and $x_i$(or $\bar{s}^2_i$) transform as,
\be
g_i &\rightarrow &{1+\sum ^n_{j=i} x_j  \over 1+\sum ^n_{j=i+1} x_j}, \\ \nonumber
F_i &\rightarrow &{ (1+x_i)(1+\sum ^n_{j=i+1} x_j) \over 1+\sum ^n_{j=i} x_j}, \\ \nonumber
x_i &\rightarrow &{x_i(1+\sum^n_{j=i} x_j)  \over  (1+\sum^n_{j=i+1} x_j)(1+\sum^n_{j=1} x_j)}.  
\ee
We also find that the exponent of $(1+\sum^n_{j=i+1}x_j)$ is given by $$({h+c_i+a_i+h-c_i+a_i \over 2}+b_i)-({h+c_{i-1}+a_{i-1}+h-c_{i-1}+a_{i-1} \over 2}+b_{i-1}),$$ and this vanishes when we use the definition of $a_i$ and $b_i$ from \eqref{aidef} and \eqref{bidef}. Hence all the terms of the form $(1+\sum_{j=i+1}^n x_j)$ do not have any contribution to the exponent. Finally we are left with a very simple integral given as
\be
\M(k_i,c_i)=\prod_{i=1}^{n} \int {ds_i \over s_i} s^{h+c_i+a_i}_i (1+s^2_i)^{d_i} \int {dx_i \over x_i} x^{h-c_i+a_i \over 2}_i (1+x_i)^{-{h+c_i+a_i \over 2}} (1+\sum^n_{j=1} x_j)^{q},
\label{mellinintegrand}
\ee
where $q={1 \over 2} (\sum c_i -(n-2)h)$. 

The $s_i$ integrals give the Gamma functions,
\be
 \prod_{i=1}^{n} \int {ds_i \over s_i} s^{h+c_i+a_i}_i (1+s^2_i)^{d_i}&=& \prod_{i=1}^{n} {\Gamma({h+c_i+a_i  \over 2}) \Gamma({ \Delta_{B_i}-c_i-h  \over 2}) \over \Gamma({ \Delta_{B_i} - a_i \over 2})},
 \label{siint}
\ee
where we have used the fact that $k^2_i = - \Delta_i$  and also the definition of $a_i$ and $d_i$ from \eqref{aidef} and \eqref{didef}to get the final form of the result.

To perform  the $x_i$ integrals, we will do a series expansion of the factor $ (1+\sum^n_{j=1} x_j)^{q} $ in (\ref{mellinintegrand}) as,
\be
	(1+\sum^n_{j=1} x_j)^{q} &=& \!\!\!\sum_{m_1,\ldots, m_n=0}^{\infty} \prod_{k=1}^n(-q\!+\sum_{j=1}^{{k-1}} m_j)_{m_k}
	\prod_{k=1}^n\frac{(-x_k)^{m_k}}{m_k!} \\ \nonumber 
	&=& \!\!\!\sum_{m_1,\ldots, m_n=0}^{\infty} \prod_{k=1}^n(-q)_{\sum_i m_i}
	\prod_{k=1}^n\frac{(-x_k)^{m_k}}{m_k!}.
\ee
Then the $x_i$ integrals can be performed easily, which leads to 
\beqa
&\prod_{i=1}^{n} &\int {dx_i \over x_i} x^{h-c_i+a_i \over 2}_i (1+x_i)^{-{h+c_i+a_i \over 2}} (1+\sum^n_{j=1} x_j)^{q} \nonumber\\
 &=&  F_A^{(n)}\left(-q,\left\{{h-c_1+a_1 \over 2},\ldots,{h-c_n+a_n \over 2}\right\},\left\{1-c_1,\ldots,1-c_n\right\};1,\ldots,1\right)\nn
 &\times & \prod_{i=1}^{n} {\Gamma(c_i) \Gamma({h-c_i+a_i \over 2}) \over \Gamma({h+c_i+a_i \over 2})}.
 \label{xiint}
\eeqa
So the Mellin integrand \eqref{mellinintegrand} is now given by the product of \eqref{siint} and \eqref{xiint}.

We can now do the final integration over the $c$ variables to get the Mellin amplitude,
\be \label{cintegration}
M(k_i)=\int^{+i \infty}_{-i \infty} \biggl(\prod^n_{i=1}{dc_i \over 2 \pi i} f_{\Delta_i}(c_i) \Gamma(\frac{ \Delta_{B_i}+c_i-h}{2}) \biggr)\Gamma(\frac{(n-2)h-\sum_{i=1}^n c_i}{2}) \M(k_i,c_i).
\ee 
As pointed out in~\cite{Paulos:2011ie}, we can do this integral by determining the poles in the kinematics, namely, the $a_i$'s and their corresponding residues. They can be determined by pinching of the contour by two poles, $c_i = \pm (\Delta_i - h)$ from $f_{\Delta_i}(c_i)$ and $c_i = a_i+h+2n_i$ from $\Gamma({h-c_i+a_i \over 2})$ with positive integer $n_i$, for each $c_i$ integration. The above mentioned residues can be cast in a simple form and we can write the full result for~\eqref{cintegration} in the following form,
\be \label{residue}
M(k_i)=\sum^{\infty}_{n_1,\dots,n_n=0} (\prod^n_{i=1}  \mathcal{P}_i) V^{\Delta_1,\dots,\Delta_{n-1},\Delta_{n_1}}_{[0,\dots,0,n_1]} \dots V^{\Delta_{(n-1)^2+1},\dots,\Delta_{n(n-1)},\Delta_{n_n}}_{[0,\dots,0,n_{n}]}
V^{\Delta_{n_1},\dots,\Delta_{n_n}}_{[n_1,\dots,n_{n}]},
\ee 
where the simple poles in $a_i$ can be read off from the  terms, $\frac{1}{a_i+(\Delta_{n_i}+2 n_i)}$, appearing in $ \mathcal{P}_i $.

One may worry about other possible poles, including the poles $\pm c_i = \Delta_{B_i} - h + 2m$ from $\Gamma({\Delta_{B_i} \pm c_i - h \over 2})$, and the pole from $\Gamma({(n-2)h - \sum_i c_i  \over 2})$. Firstly the pole from $\Gamma({(n-2)h - \sum_i c_i  \over 2})$ is canceled by $(-q)_{\sum n_i}$ in the Lauricella function, and as for the other pole, we note that after pinching off $ c_i =- (\Delta_{B_i} - h + 2m)$ with $c_i = a_i+h+2n_i$, this pole is canceled out by $\Gamma({\Delta_{B_i} + a_i \over 2})$ in $I_s(c_i)$. 

Furthermore, it has been argued in \cite{Fitzpatrick:2011ia} that the correlation function in Mellin space has good behavior at large $a_i$, and poles and the corresponding residues are enough to determine the whole function, so (\ref{residue}) is the complete result of the integral (\ref{cintegration}), and it leads to the Feynman rules stated earlier in section $2$.

For a $\phi^n$ theory, there are also vertices with less than $n$ off-shell legs. In fact one can have vertices with $n,(n\!-\!1),\dots,1$ and $0$ off-shell legs. We can obtain the results of these cases from the vertex with a maximal number of off-shell legs in Fig.~\ref{figphin} by taking some of $B_i$'s to be a single leg connecting directly to the boundary.  The result of this Witten diagram can be obtained by simply removing $s_n$ and $\bar{s}_n$ and noticing that for a single leg on the boundary we have $(t_n P_n)^2=0$. If we take $m$ out of $n$ $B_i$'s to be single legs, the result is actually in the same form of the $\phi^{n-m}$ theory, as one would have expected.

\begin{figure}[h]
  \begin{center}
\includegraphics[scale=0.6]{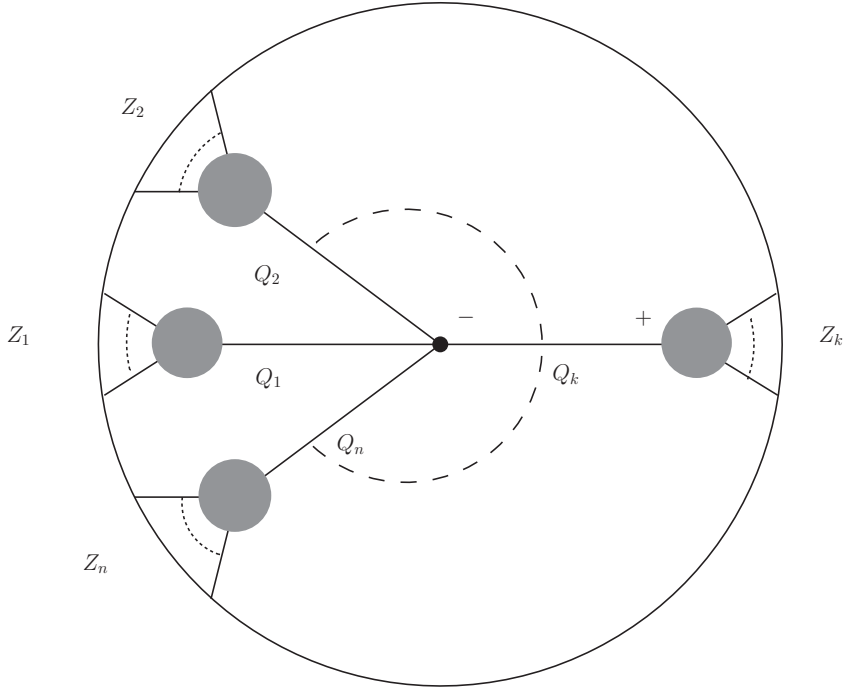}
\end{center}
\caption{ A vertex with all off-shell legs in arbitrary Witten diagram.}
  \label{figphinZ}
\end{figure}


\subsection{General case}

Let us consider the most general case of a maximal off-shell vertex in an arbitrary Witten diagram in a scalar \pn $~$ theory, as in Fig.~\ref{figphinZ}. Here, the off-shell leg $ Q_k $, is connected to the set of on-shell fields in the block named $ Z_k $ with $ k=1, \ldots, n $ via many propagators and vertices. All these intermediate propagators and vertices are collectively denoted by the blob attached to $ Q_k $.
 We also assume that we had already done the $ Q $ integrations for all the propagators inside each blob connected to the block $ Z_k $ and the contribution from this to the exponent in the integrand of the $ s, \bar s, {\rm and} ~c $ integrals is labeled  as $ \Bz_k$. We note that this quantity depends on the variables $ s, \bar s $ associated with all the propagators in the blob of the $ k_{th} $ block $ Z_k $ and all the $ t_kP_k $'s associated with the on-shell fields contained in this block,\footnote{We will label all the variables associated with the propagators in the blob with a primed index. } but most importantly it does not contain the variables associated with the propagator $ Q_k $ i.e. $s_k$ and $ \bar s_k $.

We note that Fig.~\ref{figphin} is a special case of Fig.~\ref{figphinZ} if the $ k_{th} $ blob contains only one maximal vertex of $ \phi^n $  theory with $ (n\!-\!1) $ on-shell legs and then $ \Bz_k \equiv B_k $. 

Now it is obvious that just as in the previous section, the contribution to the Mellin exponent after doing the integrations over $ Q_1 $ to $ Q_{n} $ can be written as,
 \be 
E_{Q_n}=\sum_{i'}(\D_{i'})^2 + \sum^n_{l=1}\big( \bar{s}_l \Y^Z_{l-1}+ s_l \Bz_l \big)^2, 
\label{eqgen}
\ee
where
\be
\Y^Z_{n}&=&\sum^n_{i=1} {(\prod_{k=1}^{n}g_k)\over g_i} s_i\bar{s}_i\Bz_i
\label{genericrecursion}
\ee
and $\mathcal{D}_{i'} $ is defined as follows. Since the $ Q_1 $ to $ Q_n $ integrations affect the form of $ \mathcal{B}_{i}^Z $, obtained from the previous $Q$ integrations,  at each step of these integrations we will get some complicated functions which we denote as $\mathcal{D}_{i'} $. We note that these do not have any dependence on the $ s $ and $ \bar s $ variables associated with the maximal vertex and are not of any interest for the remaining calculation\footnote{In Appendix.~B we do a specific example of a general Witten diagram  and there we also give an explicit form of these $\mathcal{D}_{i'}$'s for that special case. }. The other definitions being the same as in (\ref{recursion}). Even though, it seems that this situation is much more complicated than before we note that none of the $ \Bz_k $'s depend on the $ s_k $ and $\bar s_k $'s associated with the vertex under consideration and moreover the  $ \Bz_k $'s are also linear in the $ t_kP_k \in Z_k $ and hence we can apply Symanzik star formula, to perform the $ t_i $ integrals, as before by expanding the square.
Now, let us focus on the second sum in (\ref{eqgen}) and since it has the same form as (\ref{phin1}), the analysis is similar to the one in the previous section but with a few added subtleties. In particular, when applying Symanzik star formula, the contributions from the $\Bz_i \cdot \Bz_j$ term will be same as that of the $ B_i\cdot B_j $ term before, with $i \neq j$, but the contributions from $\Bz_i \cdot \Bz_i$ would be different from the analogous term in the previous section.\footnote{ The reason is that some terms in $\Bz_i \cdot \Bz_i$ could mix the contributions from the first sum, $\sum_{i'}(\D_{i'})^2$, in (\ref{eqgen}), however there is no such kind of mixing for the term of the form $\Bz_i \cdot \Bz_j$, because there cannot be any term in $\sum_{i'}(\D_{i'})^2$ to have this form, since the $i$th blob had never talked to $j$th blob before.} 

So all of the analysis in the previous section still holds as we can isolate the integrations for this particular vertex and write the Mellin integrand as,
\be \label{vergen1}
\mathcal{M}(k_j)&=&\int ds \mathcal{V}(s) \int \prod^n_{j=1} {ds_j \over s_j} {d\bar{s}_j \over \bar{s}_j} \, s_j^{h+c_j+a_j} \bar{s}^{h-c_j+a_j}_j g_j^{-b_j} \prod^{n}_{j=1}H_j(s^2_j F_j, s),
\ee
where we denote $\int ds \mathcal{V}(s)$ as the integrals irrelevant to the vertex. Even though $H_i(s^2_i F_i, s)$ can be a complicated function, for the $s$ and $\bar s$ relevant to the vertex we are interested in, they are always of the form $s^2_j F_j$. So we can rescale $s^2_i$ by a factor of $F_i$ for $i=1,2, \dots, n$,\footnote{Note $F_n=1$, so when $j=n$ there is no rescaling.} and after rescaling, $H_i(s^2_i F_i, s)$ will be included in the irrelevant integral $\int ds\mathcal{V}(s)$. So at the end we arrive at the integral related to the vertex we are interested in i.e. the $ \bar s $ dependent part
\be \label{vergen2}
\mathcal{M}(k_i)\big|_{\bar s}&=&\int \prod^n_{i=1}  {d\bar{s}_i \over \bar{s}_i} \, \bar{s}^{h-c_i+a_i}_i g_i^{-b_i} \prod^{n}_{i=1} F^{\frac{-h+c_i+a_i}{2}}_i,
\ee
which is exactly the same as the $ \bar s $  part of the integral in (\ref{phin3}) and hence gives the same form of the maximal off-shell vertex. 
 
\section{Flat space limit}
In this section we consider the flat space limit of the Mellin space Feynman rules. We will show that these rules give rise to the usual Feynman rules for scattering amplitudes in the flat space limit. This limit can also be considered as a consistency check of the AdS Feynman rules. The flat space limit corresponds to the large $\delta_{ij}$ behavior of the Mellin amplitudes. As had been discussed in \cite{Penedones:2010ue} and \cite{Fitzpatrick:2011ia}, in this limit the Mellin amplitudes are related to the S-matrix in flat space by the following relation\footnote{Notice it is slightly different from Eq.~$(127)$ in~\cite{Fitzpatrick:2011ia}, because we define the Mellin amplitudes by a different normalization factor.}
\beq
M(\delta_{ij}) \approx  
\ \int_0^\infty d\beta \,\beta^{\frac{1}{2}\sum \Delta_i -h-1} e^{-\beta}\,
T(p_i\cdot p_j=2 \beta \delta_{ij})\ ,\ \ \ \ \ \ \ \ 
\delta_{ij} \gg 1\ ,
\label{FSlimit}
\eeq
where  $ T(p_i\cdot p_j=2 \beta \delta_{ij}) $ is the flat space S-matrix as a function of the kinematic invariants $ p_i\cdot p_j $ and $ \beta $ is an integration parameter. We will study the case of large $\delta_{ij}$ limit with $\Delta_i$ fixed. In order to confirm that the AdS Feynman rules indeed reduce to the usual flat space Feynman rules of $\phi^n$ theory in this limit, we only need to show that
\be \label{flatlimit}
\sum_{\{n_i\}} M(n_1, \dots, n_s) ={(-1)^s \over 2^s} \Gamma({1\over 2} \sum \Delta_i - h -s), 
\ee
where $\sum_{\{n_i\}} M(n_1, \dots, n_s)$ related to the Mellin amplitude by
\be
M(\delta_{ij}) \approx \sum_{\{n_i\}} M(n_1, \dots, n_s) \prod^s_{i=1} {1 \over k^2_i}, 
\ee
where $1 \over k^2_i$ is the propagator, $s$ is the number of propagators, and the summation over $\{n_i\}$ become clear shortly. The above equation~\eqref{flatlimit} follows directly from~(\ref{FSlimit}) by using the definition of the flat space massless scalar scattering amplitudes.

 We will use the AdS Feynman rules~\eqref{lauricella} to compute the left hand side of~(\ref{FSlimit}). As in the case of $\phi^3$ theory~\cite{Fitzpatrick:2011ia}, we can always start from the bulk propagators closer to the external legs in the Witten diagram and perform the sum over $\{ n_i \}$ recursively. To do so for a general $\phi^n$ theory, we need the following identity, which will be proved shortly,
\be \label{identity2}
&&\sum^{\infty}_{n_{I_1},n_{I_2}, \dots, n_{I_{m}}=0}  { V^{(1)}(n_{I_1}) V^{(1)}(n_{I_2})\dots V^{(1)}(n_{I_m})V^{(m+1)}(n_{I_1},n_{I_2}, \dots, n_{I_m},n_o)  \over P_{n_{I_1}} P_{n_{I_2}} \dots P_{n_{I_m}} } \\ \nonumber 
&&= {(-1)^m\over 2^m} ((m+1) - {\sum \Delta_i \over 2}+ \dd {n_o})_{n_o} \Gamma({\sum \Delta_i \over 2} - h - m),
\ee
where $P_n \equiv -2 n! \Gamma(1+\Delta_{n} + n-h)$, and $V^{(1)}(n_I)$ denotes the vertex with one off-shell leg where this leg is labeled as $n_I$, and we follow a similar logic to define $V^{(m+1)}(n_{I_1},n_{I_2}, \dots, n_{I_m},n_o)$, which denotes the vertex with $ (m+1) $ off-shell legs. Finally, the summation in $\sum \Delta_i$ indicates the sum over all the external on-shell legs. The identity~(\ref{identity2}) can be proved by performing the summation in the following order: first we sum over $l_1$, the summation variable in the Lauricella function $F_A^{(n)}$ then the corresponding $n_{I_1}$, next we do the sum over $l_2$ then $n_{I_2}$ and so on. At each step of the sum we can apply the identity,
\beq
\sum^{\infty}_{n=0} {(a)_n (b)_n \over n! (c)_n} = {(c-b)_{-a} \over (c)_{-a}}.
\eeq

We note here that $((m+1) - {\sum \Delta_i \over 2}+ \dd {n_o})_{n_o}$ in~(\ref{identity2}) has the same dependence on $n_o$ as $V^{(1)}(n)$ does. So the summation on $n_o$ in a general Witten diagram can be done by applying the above identities again. At the end of the day, we will be left with a sum involving only factors of the form of $V^{(1)}(n)$. It can be shown that the answer for the final sum of any Witten diagram is indeed given as Eq.~(\ref{flatlimit}). 


\begin{acknowledgments}

We are grateful to thank Jared Kaplan, Miguel Paulos, Marcus Spradlin and Gabriele Travaglini
for very useful conversations. This work is supported in part by
by the US Department of Energy under contracts
DE-FG02-91ER40688 (Task A) and DE-FG02-11ER41742 (Early Career Award), 
the US National Science Foundation under grant
PHY-0643150 (PECASE) and Sloan Research Fellowship.
C.W. would like to acknowledge the support of the STFC Standard Grant ST/J000469/1 ``String
Theory, Gauge Theory and Duality".

\end{acknowledgments}
\appendix

\section{Useful integrals}
Here we list some formulas, which have been extensively used in this paper. For more details about these formulas, see \cite{Symanzik:1972wj,Penedones:2010ue}.

\subsection*{$X$ integral formula}
	\beq
	\int_{0}^{+\infty} \prod_{i}\left(\frac{\ud t_i}{t_i} t^{\alpha_i}\right) \iads \ud X\, e^{2 T\cdot X}
	=
	\pi^h \gt{\sum_i \alpha_i-2h} 
	\int_{0}^{+\infty} \prod_{i}\left(\frac{\ud t_i}{t_i}  t^{\alpha_i}\right) e^{T^2}, \label{Xint}
	\eeq
	where $T= \sum_i t_iP_i$.
	\subsection*{$Q_i$ integral formula}

	\beq
	\int_{0}^{+\infty} \frac{\ud s}{s} \frac{\ud \bar s}{\bar s} 	s^{h+c}s^{h-c}\,
	\ib \ud Q\, e^{2 T\cdot Q}
	=2\pi^h \int_{0}^{+\infty} \frac{\ud s}{s} \frac{\ud \bar s}{\bar s} s^{h+c}s^{h-c} e^{T^2},	\label{boundint}
	\eeq
	where $T= (sX + \bar{s}Y)$.
\subsection*{$ t $ integral formula}
This is also called  the Symanzik star integration formula  where we consider a set of $n$ points in Euclidean space $x_i$ and their differences $x_i-x_j$. In the embedding formalism we have $P_{ij}\equiv -2 P_i\cdot P_j=(x_i-x_j)^2$. Then Symanzik's formula is:
	\beqa
	 \int_{0}^{+\infty}\!\! \left(\prod_{i=1}^n \frac{\ud t_i}{t_i} t^{\Delta_i}\right) e^{-(\sum\limits_{1\leq i<j \leq n} t_i t_j\, P_{ij})}=
		\frac{\pi^h/2}{(2\pi i)^{\frac 12 n(n-3)}}\int \ud \delta_{ij}\! \prod_{1\leq i<j \leq n}\!\!\Gamma(\delta_{ij})\,(P_{ij})^{-\delta_{ij}} \label{symanzikint}
	\eeqa
where the integration is over $ n(n-3)/2 $ variables and the integration paths are chosen parallel to the imaginary axis, with real parts such that the real parts of the arguments of the gamma functions are positive.

\section{Example: $ 9 $-points in $ \phi^3 $ theory}

\begin{figure}[h]
  \begin{center}
\includegraphics[scale=0.6]{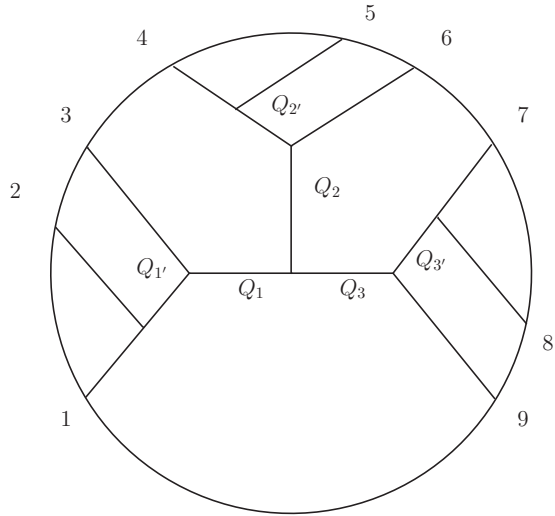}
\end{center}
\caption{ A vertex with all off-shell legs in $9$-point amplitude in $\phi^3$ theory }
  \label{9ptWD}
\end{figure}

In order to illustrate the general strategy of isolating the maximal vertex in any arbitrary Witten diagram, as discussed in section $ 3.3 $, here we will study a specific example, that of a $ 9$-point Witten diagram in  $ \phi^3 $ theory and we will write down the results as a special case of  \eqref{vergen1} and \eqref{vergen2}. We have the $ Q $ variables labeled as $  Q_{1'}, Q_{2'}, Q_{3'}, Q_1, Q_2, Q_{3} $ and then we will integrate them out in that particular order. We note here that only the variables $ Q_{1}, Q_{2}, Q_{3}  $ are relevant for the vertex under consideration. After doing all the $ Q $ integrals we will get the exponent as before  given by \eqref{eqgen},
\be 
E_{Q_3}=\sum_{i=1}^3(\D_{i'})^2 + \sum^3_{l=1}\big( \bar{s}_l \Y^Z_{l-1}+ s_l \Bz_l \big)^2, 
\ee
where the term in the first bracket is given by
\beqa
\D_{1'} &=& P_{3} t_{3} \bar{s}_{1'}+s_{1'} \left(P_{1} t_{1}+P_{2}
   t_{2}\right),\nn
  \D_{2'} &=&  P_{6} t_{6} \bar{s}_{2'}+s_{2'} \left(P_{4} t_{4}+P_{5}
   t_{5}\right),\nn
   \D_{3'} &=& P_9 t_9 \bar{s}_{3'}+s_{3'} \left(P_7 t_7+P_8
   t_8\right),
   \label{curlyD}
 \eeqa
 and these terms do not contain any of the variables relevant to the vertex we are interested in and hence we will not consider them further.  Now let us focus on the term inside the second bracket where the relevant functions $ \Bz $ and $ \Y^Z $, defined by \eqref{genericrecursion}, are given by,
\beqa
\Bz_{1} &=& \mathcal{D}_{1'} \bar{s}_{1'}+P_{3} t_{3},\nn
\Bz_{2} &=& \mathcal{D}_{2'} \bar{s}_{2'}+P_{6} t_{6},\nn
\Bz_{3} &=& \mathcal{D}_{3'} \bar{s}_{3'}+P_9 t_9
\label{Bz}
\eeqa
and
\beqa
\Y^Z_0 &=& 0,\nn
\Y^Z_{1} &=& s_{1} \Bz_{1} \bar{s}_{1}, \nn
\Y^Z_{2} &=& s_{2} \Bz_{2}
   \left(\bar{s}_{1}^2+1\right)
   \bar{s}_{2}+s_{1} \Bz_{1} \bar{s}_{1}
   \left(\left(\bar{s}_{1}^2+1\right)
   \bar{s}_{2}^2+1\right).
   \eeqa
 The next step is to expand the second term of $ E_{Q_3} $ like before and after doing all the $ t_i $ integrals we  get the  form of the integrand as in \eqref{vergen1},
 \be
\mathcal{M}(k_j)&=&\int ds \mathcal{V}(s) \int \prod^3_{j=1} {ds_j \over s_j} {d\bar{s}_j \over \bar{s}_j} \, s_j^{h+c_j+a_j} \bar{s}^{h-c_j+a_j}_j g_j^{-b_j} \prod^{3}_{j=1}H_j(s^2_j F_j, s),
\ee
where $ g $ and $ F $ are defined by \eqref{recursion} and \eqref{Aexpand0} and $ a_j, b_j $ are defined by \eqref{aidef} and \eqref{bidef} respectively. Moreover, the part which is irrelevant to the maximal vertex is given by 
\beqa
\mathcal{V}(s) &=& \prod^{3'}_{j=1'} {ds_j \over s_j} s_j^{h+c_j+a'_j} \bar s_j^{h-c_j+a'_j}\left(\bar{s}_{1'}^2+1\right){}^{-k_3
   \left(k_1+k_2+k_3\right)}\nn
  &\times & \left(\bar{s}_{2'}^2+1\right){}^{-k_6
   \left(k_4+k_5+k_6\right)}
   \left(\bar{s}_{3'}^2+1\right){}^{-k_9
   \left(k_7+k_8+k_9\right)},
   \eeqa   
  where,
  \beqa
  a'_{1'} &=& -(k_1+k_2)^2,\nn 
    a'_{2'} &=& -(k_4+k_5)^2,\nn 
  a'_{3'} &=& -(k_7+k_8)^2. 
\eeqa
 The complicated $ H $ function which would eventually give the terms relevant for the vertex, is given by,
\be
H_1(s^2_1 F_1, s) &=& \left(F_{1} s_{1}^2
   \left(\bar{s}_{1'}^2+1\right)+1\right){}^{\left(k_1+k_2\right) k_3} \left(s_{1'}^2
   \left(F_{1} s_{1}^2
   \bar{s}_{1'}^2+1\right)+1\right){}^{k_1 k_2},\nn
H_2(s^2_2 F_2, s)&=&   \left(F_{2} s_{2}^2
   \left(\bar{s}_{2'}^2+1\right)+1\right){}^{\left(k_4+k_5\right) k_6} \left(s_{2'}^2
   \left(F_{2} s_{2}^2
   \bar{s}_{2'}^2+1\right)+1\right){}^{k_4 k_5},\nn
H_3(s^2_3 F_3, s)&= &   \left(F_{3} s_{3}^2
   \left(\bar{s}_{3'}^2+1\right)+1\right){}^{\left(k_7+k_8\right) k_9} \left(s_{3'}^2
   \left(F_{3} s_{3}^2
   \bar{s}_{3'}^2+1\right)+1\right){}^{k_7 k_8}.
   \label{Hcomplicated}
\ee
 Now, as before, we can use the the transformation 
 \beqa
 s_i^2 \rightarrow \frac{s_i^2}{F_i}
 \eeqa
 and we will get the required form of the $\bar s $ integral part as in \eqref{vergen2}, to be
 \be 
\mathcal{M}(k_i)\big|_{\bar s}&=&\int \prod^3_{i=1} {d\bar{s}_i \over \bar{s}_i} \, \bar{s}^{h-c_i+a_i}_i g_i^{-b_i} F^{\frac{-h+c_i+a_i}{2}}_i.
\ee  
From \eqref{Hcomplicated} we also see that after the rescaling of $ s^2_i $'s we are left with a term of the form ,
\beqa
\mathcal{V}'(s) &=& \left(s_{1}^2
   \left(\bar{s}_{1'}^2+1\right)+1\right){}^{\left(k_1+k_2\right) k_3} \left(s_{1'}^2
   \left(s_{1}^2
   \bar{s}_{1'}^2+1\right)+1\right){}^{k_1 k_2}\nn
&\times &   \left( s_{2}^2
   \left(\bar{s}_{2'}^2+1\right)+1\right){}^{\left(k_4+k_5\right) k_6} \left(s_{2'}^2
   \left(s_{2}^2
   \bar{s}_{2'}^2+1\right)+1\right){}^{k_4 k_5}\nn
&\times &   \left(s_{3}^2
   \left(\bar{s}_{3'}^2+1\right)+1\right){}^{\left(k_7+k_8\right) k_9} \left(s_{3'}^2
   \left(s_{3}^2
   \bar{s}_{3'}^2+1\right)+1\right){}^{k_7 k_8},
\eeqa
which we note is independent of the $ \bar s $ variables related to the maximal vertex and hence irrelevant.


\newpage
\begin{thebibliography}{99}

\bibitem{Maldacena:1997re}
J.~M. Maldacena, ``{The Large N limit of superconformal field theories and
  supergravity},'' {{\em Adv.Theor.Math.Phys.} {\bfseries 2} (1998)
  231--252}, {{\ttfamily
  arXiv:hep-th/9711200 [hep-th]}}.

\bibitem{Witten:1998qj}
E.~Witten, ``{Anti-de Sitter space and holography},'' {\em
  Adv.Theor.Math.Phys.} {\bfseries 2} (1998) 253--291,
  {{\ttfamily arXiv:hep-th/9802150
  [hep-th]}}.

\bibitem{Gubser:1998bc}
S.~Gubser, I.~R. Klebanov, and A.~M. Polyakov, ``{Gauge theory correlators from
  noncritical string theory},''
  {{\em Phys.Lett.}
  {\bfseries B428} (1998) 105--114},
 {{\ttfamily arXiv:hep-th/9802109
  [hep-th]}}.

\bibitem{LiuTseytlin}
H.~Liu and A.~A. Tseytlin, ``{On four-point functions in the CFT/AdS
  correspondence},''{{\em
  Phys. Rev.} {\bfseries D59} (1999) 086002},
{{\ttfamily arXiv:hep-th/9807097}}.

\bibitem{hep-th/9811152} 
  H.~Liu,
  ``Scattering in anti-de Sitter space and operator product expansion,''
  Phys.\ Rev.\ D\ {\bf 60}, 106005  (1999)
 {{\ttfamily  arXiv:hep-th/9811152}}.


\bibitem{D'Hoker:1998mz}
E.~D'Hoker and D.~Z. Freedman, ``{General scalar exchange in AdS(d+1)},''
  {{\em Nucl. Phys.}
  {\bfseries B550} (1999) 261--288},
{{\ttfamily arXiv:hep-th/9811257}}.

\bibitem{Freedman:1998bj}
D.~Z. Freedman, S.~D. Mathur, A.~Matusis, and L.~Rastelli, ``{Comments on
  4-point functions in the CFT/AdS correspondence},''
  {{\em Phys. Lett.}
  {\bfseries B452} (1999) 61--68},
{{\ttfamily arXiv:hep-th/9808006}}.

\bibitem{Freedman:1998tz}
D.~Z. Freedman, S.~D. Mathur, A.~Matusis, and L.~Rastelli, ``{Correlation
  functions in the CFT($d$)/AdS($d+1$) correspondence},''
 {{\em Nucl. Phys.}
  {\bfseries B546} (1999) 96--118},
{{\ttfamily arXiv:hep-th/9804058}}.

\bibitem{D'Hoker}
E.~D'Hoker, D.~Z. Freedman, S.~D. Mathur, A.~Matusis, and L.~Rastelli,
  ``{Graviton exchange and complete 4-point functions in the AdS/CFT
  correspondence},''
  {{\em Nucl. Phys.}
  {\bfseries B562} (1999) 353--394},
{{\ttfamily arXiv:hep-th/9903196}}.

\bibitem{Arutyunov:2000py}
G.~Arutyunov and S.~Frolov, ``{Four-point functions of lowest weight CPOs in N
  = 4 SYM(4) in supergravity approximation},''
 {{\em Phys. Rev.}
  {\bfseries D62} (2000) 064016},
{{\ttfamily arXiv:hep-th/0002170}}.

\bibitem{Arutyunov:2002fh}
G.~Arutyunov, F.~A. Dolan, H.~Osborn, and E.~Sokatchev, ``{Correlation
  functions and massive Kaluza-Klein modes in the AdS/CFT correspondence},''
 {{\em Nucl. Phys.}
  {\bfseries B665} (2003) 273--324},
{{\ttfamily arXiv:hep-th/0212116}}.

\bibitem{Arutyunov:2003ae}
G.~Arutyunov and E.~Sokatchev, ``{On a large N degeneracy in N = 4 SYM and the
  AdS/CFT correspondence},''
  {{\em Nucl. Phys.}
  {\bfseries B663} (2003) 163--196},
{{\ttfamily arXiv:hep-th/0301058}}.

\bibitem{Berdichevsky:2007xd}
L.~Berdichevsky and P.~Naaijkens, ``{Four-point functions of different-weight
  operators in the AdS/CFT correspondence},''
  {{\em JHEP} {\bfseries
  01} (2008) 071},
{{\ttfamily arXiv:0709.1365 [hep-th]}}.

\bibitem{Uruchurtu:2008kp}
L.~I. Uruchurtu, ``{Four-point correlators with higher weight superconformal
  primaries in the AdS/CFT Correspondence},''
  {{\em JHEP} {\bfseries
  03} (2009) 133},
{{\ttfamily arXiv:0811.2320 [hep-th]}}.

\bibitem{Buchbinder:2010vw}
E.~I. Buchbinder and A.~A. Tseytlin, ``{On semiclassical approximation for
  correlators of closed string vertex operators in AdS/CFT},''
  {{\em JHEP} {\bfseries 08}
  (2010) 057},
{{\ttfamily arXiv:1005.4516 [hep-th]}}.

\bibitem{Uruchurtu:2011wh}
L.~I. Uruchurtu, ``{Next-next-to-extremal Four Point Functions of N=4 1/2 BPS
  Operators in the AdS/CFT Correspondence},''
{{\ttfamily arXiv:1106.0630 [hep-th]}}.

\bibitem{Dolan:2006ec}
F.~A. Dolan, M.~Nirschl, and H.~Osborn, ``{Conjectures for large N N = 4
  superconformal chiral primary four point functions},''
  {{\em Nucl. Phys.}
  {\bfseries B749} (2006) 109--152},
{{\ttfamily arXiv:hep-th/0601148}}.

\bibitem{Howtozintegrals}
E.~D'Hoker, D.~Z. Freedman, and L.~Rastelli, ``{AdS/CFT 4-point functions: How
  to succeed at z-integrals without really trying},''
  {{\em Nucl. Phys.}
  {\bfseries B562} (1999) 395--411},
{{\ttfamily arXiv:hep-th/9905049}}.



\bibitem{Raju1}
S.~Raju, ``{BCFW for Witten Diagrams},''
 {{\em Phys.Rev.Lett.}
  {\bfseries 106} (2011) 091601},
  {{\ttfamily arXiv:1011.0780 [hep-th]}}.

\bibitem{Raju2}
S.~Raju, ``{Recursion Relations for AdS/CFT Correlators},''
  {{\em Phys. Rev.}
  {\bfseries D83} (2011) 126002},
{{\ttfamily arXiv:1102.4724 [hep-th]}}.


\bibitem{Mack:2009mi} 
  G.~Mack,
  ``D-independent representation of Conformal Field Theories in D dimensions via transformation to auxiliary Dual Resonance Models. Scalar amplitudes,''
 {{\ttfamily  arXiv:0907.2407 [hep-th]}}.


\bibitem{Mack:2009gy} 
  G.~Mack,
  ``D-dimensional Conformal Field Theories with anomalous dimensions as Dual Resonance Models,''
  {{\ttfamily arXiv:0909.1024 [hep-th]}}.


\bibitem{Penedones:2010ue} 
  J.~Penedones,
  ``Writing CFT correlation functions as AdS scattering amplitudes,''
  JHEP\ {\bf 1103}, 025  (2011)
 {{\ttfamily  arXiv:1011.1485 [hep-th]}}.


\bibitem{Paulos:2011ie}
  M.~F.~Paulos,
  ``Towards Feynman rules for Mellin amplitudes,''
   {{\ttfamily  arXiv:1107.1504 [hep-th]}}.

\bibitem{Fitzpatrick:2011ia}
  A.~L.~Fitzpatrick, J.~Kaplan, J.~Penedones, S.~Raju and B.~C.~van Rees,
  ``A Natural Language for AdS/CFT Correlators,''
  {{\ttfamily arXiv:1107.1499 [hep-th]}}.


\bibitem{Costa:2011mg}
  M.~S.~Costa, J.~Penedones, D.~Poland and S.~Rychkov,
  ``Spinning Conformal Correlators,''
 {{\em JHEP} {\bf 1111}, 071 (2011)}
  {{\ttfamily arXiv:1107.3554 [hep-th]}}.
  
\bibitem{Costa:2011dw}
  M.~S.~Costa, J.~Penedones, D.~Poland and S.~Rychkov,
  ``Spinning Conformal Blocks,''
  {{\ttfamily arXiv:1109.6321 [hep-th]}}.
\bibitem{arXiv:1102.0577} 
  I.~Balitsky,
  ``Mellin representation of the graviton bulk-to-bulk propagator in AdS,''
  {{\em Phys.\ Rev.\ D}\ {\bf 83}, 087901  (2011)}
    {{\ttfamily arXiv:1102.0577 [hep-th]]}}.


\bibitem{susskind}
L.~Susskind, ``{Holography in the flat space limit},''
{{\ttfamily arXiv:hep-th/9901079}}.

\bibitem{polchinski}
J.~Polchinski, ``{S-matrices from AdS spacetime},''
{{\ttfamily arXiv:hep-th/9901076}}.

\bibitem{GGP}
M.~Gary, S.~B. Giddings, and J.~Penedones, ``{Local bulk S-matrix elements and
  CFT singularities},''
 {{\em Phys. Rev.}
  {\bfseries D80} (2009) 085005},
{{\ttfamily arXiv:0903.4437 [hep-th]}}.

\bibitem{JP}
I.~Heemskerk, J.~Penedones, J.~Polchinski, and J.~Sully, ``{Holography from
  Conformal Field Theory},''
  {{\em JHEP} {\bfseries
  10} (2009) 079},
{{\ttfamily arXiv:0907.0151 [hep-th]}}.

\bibitem{Katz}
A.~L. Fitzpatrick, E.~Katz, D.~Poland, and D.~Simmons-Duffin, ``{Effective
  Conformal Theory and the Flat-Space Limit of AdS},''
{{\ttfamily arXiv:1007.2412 [hep-th]}}.

\bibitem{TakuyaFSL}
T.~Okuda and J.~Penedones, ``{String scattering in flat space and a scaling
  limit of Yang-Mills correlators},''
{{\ttfamily arXiv:1002.2641 [hep-th]}}.

\bibitem{Fitzpatrick:2011jn}
A.~Fitzpatrick and J.~Kaplan, ``{Scattering States in AdS/CFT},''
  {{\ttfamily arXiv:1104.2597 [hep-th]}}.

\bibitem{Gary:2009mi}
M.~Gary and S.~B. Giddings, ``{The flat space S-matrix from the AdS/CFT
  correspondence?},'' {{\em
  Phys. Rev.} {\bfseries D80} (2009) 046008},
{{\ttfamily arXiv:0904.3544 [hep-th]}}.

\bibitem{GiddingsBulkLoc}
S.~B. Giddings, ``{Flat-space scattering and bulk locality in the AdS/CFT
  correspondence},'' {{\em
  Phys. Rev.} {\bfseries D61} (2000) 106008},
{{\ttfamily arXiv:hep-th/9907129}}.

\bibitem{Gary:2011kk}
M.~Gary and S.~B. Giddings, ``{Constraints on a fine-grained AdS/CFT
  correspondence},'' {{\ttfamily
  {{\ttfamily arXiv:1106.3553 [hep-th]}}}}. 



\bibitem{Symanzik:1972wj} 
  K.~Symanzik,
  ``On Calculations in conformal invariant field theories,''
  Lett.\ Nuovo Cim.\ \ {\bf 3}, 734  (1972).




\bibitem{Fitzpatricknew}
A.~Fitzpatrick and J.~Kaplan, ``{Analyticity and the Holographic S-Matrix},''
  {{\ttfamily 	arXiv:1111.6972v1 [hep-th]}}.

\end{thebibliography}
\end{document}